\documentstyle[aps]{revtex}
\newcommand{\beq}{\begin{equation}}
\newcommand{\eeq}{\end{equation}}

\newcommand{\beqa}{\begin{eqnarray}}
\newcommand{\eeqa}{\end{eqnarray}}

\newcommand{\FID}{{\mathcal{D}}}

\title{$p$-Branes Electric--Magnetic Duality and Stueckelberg/Higgs Mechanism:
       a Path--Integral Approach\footnote{Accepted for publication in
       {\bf Progr. Th. Phys.}}}

\author{Stefano Ansoldi\footnote{E-mail address: ansoldi@trieste.infn.it}}
\address{Dipartimento di Fisica Teorica \break
		 Universit\`a di Trieste, \\
		 INFN, Sezione di Trieste}

\author{Antonio Aurilia\footnote{E-mail address: aaurilia@csupomona.edu}}
\address{Department of Physics \\
		 California State Polytechnic University \break
         Pomona, CA 91768}

\author{Luca Marinatto\footnote{E-mail address: marinat@trieste.infn.it}}
\address{Dipartimento di Fisica Teorica \break
		 Universit\`a di Trieste}

\author{Euro Spallucci\footnote{E-mail address: spallucci@trieste.infn.it}}
\address{Dipartimento di Fisica Teorica \break
		 Universit\`a di Trieste, \\
		 INFN, Sezione di Trieste}

\begin{document}

\maketitle

\begin{abstract}
We study the vacuum functional for a system of $p$-branes interacting with
Maxwell fields of higher rank. This system represents a generalization of
the usual electrodynamics of point particles, with one essential
difference: namely, that the world--history of a p--brane, due to the
spatial extension of the object, may possess a physical boundary. Thus, the
objective of this study is twofold: first, we wish to exploit the breaking
of gauge invariance due to the presence of a physical boundary, in order to
generate mass as an alternative to the Higgs mechanism; second, we wish to
investigate how the new mechanism of mass generation is affected by the
duality transformation between electric and magnetic branes.\\
 The whole analysis is performed by using the path--integral method, as
opposed to the more conventional canonical approach. The advantage of the
path integral formulation is that it enables us to Fourier transform the
field strength directly, rather than the gauge potential. To our knowledge,
this {\it field strength} formulation represents a new application of the
path integral method, and leads, in a straightforward way, to the dual
representation of the vacuum functional. We find that the effect of the
dual transformation is essentially that of exchanging the role of the gauge
fields defined respectively on the `` bulk'' and ``boundary" of the
p--brane history.
\end{abstract}

\newpage

\section{Introduction}
\subsection{Synopsis and Objectives}

Electric--magnetic duality for closed $p$-branes embedded in a $D$-dimensional
target space was established many years ago\cite{nepo}, leading to the
general correspondence that the dual of a $p$-brane is a $\widetilde p$-brane
with $\widetilde p=D-p-4$. Table[\ref{table1}] illustrates this correspondence.

\begin{table}
\caption{Closed $p$-branes duality in $D=11$ spacetime dimensions.}
	\begin{tabular}{ccc}
		$p$ & $p$-brane & $\tilde p=7-p$ dual--brane
	   	\\
		\tableline
	    \\
		0 & point--particle & 7--brane
	    \\
		1 & string & 6--brane
	    \\
	    2 & membrane & 5--brane
	    \\
		3 & bag & 4--brane
	    \\
		\\
	\end{tabular}
\label{table1}
\end{table}

For instance, the brane solution of $(D=11)$ supergravity is the
``magnetic dual'' of an electric membrane, and there is a possibility that
this type of solutions may represent the basic geometric elements of a
unified theory of all fundamental interactions. Among the open questions that
such a ``final'' theory must address, a preeminent one concerns the mechanism
of mass generation in the universe. The celebrated Higgs mechanism was
invented, and successfully applied, within the framework of  local quantum
field theory, i.e., a ``low energy'' framework dealing with the interactions of
point--particles in the Standard Model. Since point--particles are currently
thought of as low energy manifestations of the underlying dynamics of strings
and higher dimensional objects, it seems pertinent to ask two fundamentally
related questions: first, \textit{what is the
``engine'' of mass production at the level of $p$-brane dynamics,} say at the
string scale of energy and beyond; and second, \textit{how is the new
mechanism of mass generation affected by the duality transformation} which
plays such a significant role in the theory of extended objects.\\
With the above questions in mind, we shall  extend the notion of
electric--magnetic
duality to the case of {\it open} $p$-branes. Apart from its intrinsic
interest as an example of duality transformation, this
extension of electric--magnetic duality will enable us to incorporate a new
mechanism of mass generation which
stems directly from the presence of a boundary and from a non-trivial
interaction between gauge fields of different
rank defined respectively on the bulk and boundary of the p--brane history.\\
To our mind, this boundary effect, and concomitant mixing of gauge fields,
represents a geometric realization of the Higgs mechanism never discussed
in the physics of point-particles for the simple reason that the
world--history of a point--particle is usually assumed to have no boundary.
On the other hand, the new mechanism of mass generation is
\textit{precisely} a boundary effect, and in order to fully appreciate its
meaning one must keep in mind the historical relationship between mass and
gauge invariance. At first glance, these two concepts are contradictory, in
the sense that mass \textit{explicitly} breaks gauge invariance. On the
other hand, mass is required for obvious experimental reasons, while gauge
invariance is the essential prerequisite for the self--consistency (for
instance, renormalizability) of any successful model of particle
interactions. In the Standard Model, mass and gauge invariance are
reconciled through the loophole of spontaneous symmetry breaking,
and the ensuing Higgs mechanism.  {\it Upon fixing a gauge, say the unitary
gauge, the role of the mass becomes manifest.}\\
 Thus, assuming that gauge invariance is just as important in the theory of
extended objects as it is for point particles, what we wish to suggest in
this paper is a geometric variation of the Higgs mechanism which is
{\it consistent} with the requirement of gauge invariance in the theory of
p--branes, both electric and magnetic, in interaction with higher rank
gauge fields.\\
On the mathematical side, the originality of our approach stems from a new
application of path--integral techniques to {\it open}
$p$-branes. A self--contained discussion of this ``sum over histories
approach'' is presented in Appendix \ref{dualityapp} which constitutes the
mathematical backbone of the paper; the method is centered on the use of a
functional representation of the Dirac--delta distribution as a tool to
implement the duality transformation directly on the gauge field strength,
rather than the gauge potential. As we shall see, this new technique has the
following desirable properties: i) it clearly separates the dynamics of the
``bulk'' from the dynamics of the ``boundary'' of the p--brane history; ii) it
illustrates how bulk and boundary fields exchange their role as a
consequence of the duality transformation; finally, iii) it is especially
advantageous in handling the Bianchi identities in higher rank gauge
theories which include both ``electric''
and ``magnetic'' objects. \\

\subsection{Conventions and Outline of Paper}

In the following analysis, we shall assume that the dimensionality of the
target spacetime $D> p+1$, in order to deal with \textit{dynamical}
extended objects. The interesting limiting case $D=p+1$, in which there are
no propagating degrees of freedom, requires a separate discussion, and can
be found in one of our recent publications\cite{bags}.\\
In order to bring out the physical content of our discussion with the bare
minimum of formalism, we have confined the bulk of our calculations to
one self--contained appendix. However, we believe that those
calculations represent, by themselves, a new and noteworthy contribution to
the literature devoted to the case of open branes, especially in view of
the increasing importance of extended objects, such as D--branes, in the
formulation of non perturbative string theory.
Furthermore, in order to avoid a cumbersome proliferation of indices, in
the main text we shall adopt the followig index--free notation:
\begin{eqnarray}
&& A^ {( p+1) } \equiv A^ {\mu_1\dots \mu_ { p+1} }\nonumber\\
&& A_ {( p+1 )} \equiv A_ {\mu_1\dots \mu_ { p+1} }\nonumber\\
&& A_ {( p+1 )} J^ { ( p+1 )}\equiv {1\over (p+1)!  } A_ {\mu_1\dots \mu_ {
p+1}
}
J ^{\mu_1\dots \mu_ { p+1} }\nonumber\\
&& \partial J^{ (p+1) }\equiv\partial_ { \mu_1 }J ^{\mu_1\dots \mu_ { p+1} }
\nonumber\\
&& d J^{ (p+1) }\equiv\partial^ {[\, \nu_1 }J ^{\mu_1\dots \mu_ { p+1}\,] }
\nonumber\\
&& d A_{ (p+1) }\equiv\partial_ {[\, \mu_1 }A _{\mu_1\dots \mu_ { p+1}\,] }
\nonumber\\
&& { 1\over \Box}  J^{ (p+1) }(x)\equiv \int dy\, G(\, x-y\, ) J^{ (p+1) }(y)
\nonumber\\
&& { }^* K^ { (p+1) } \equiv {1\over (D-p-1 ) !}
\epsilon ^{\mu_1\dots\mu_{p+1}
\mu_ {p+2}\dots\mu_d}K_{\mu_ {p+2}\dots\mu_d }\nonumber\\
&& D_ { (D-p-2) } {} ^* d I_{ (p+1) }= (-1)^ { Dp } I^ { (p+1) }{} ^* dD^ {
(D-p-2) }\nonumber\\
&& N ^ { (p+1) }= (-1)^ { D(p+1)- (p+1)^2 } {} ^*[\, {} ^*( N ^ { (p+1)
}\,)\,] .
\end{eqnarray}
For book--keeping purposes, in the above list, the upper or lower index on
the left hand side is a reminder of the actual number of indices carried by
the corresponding tensor on the right hand side.
Finally, as a further notational simplification, we shall omit the volume
of integration symbol, $d^Dx$, wherever a spacetime integration is performed.\\
With the above remarks and conventions in mind, the plan of the paper is as
follows:\\
In Section II, we discuss the Stueckelberg mechanism for open, electric,
$p$-branes. This example will clarify how mass is tied up with the
existence of a boundary, and how gauge invariance is preserved, albeit in
an extended form.\\
In Section III, we discuss the quantization condition
of electric and magnetic charges as a consequence of the arbitrariness in the
choice of the ``Dirac parent brane, '' followed by a brief introduction to
the duality symmetry
among electric and magnetic $p$-branes. \\
In Section IV, we perform a duality rotation and study the Stueckelberg
mechanism in the ``magnetic phase'' of the model.\\
Section V is devoted to a summary and discussion of the results.\\
 Appendix \ref{dualityapp} provides the details of the computations leading to
 the dual action for a system of interacting electric and magnetic
$p$-branes.\\
A set of accompanying Tables should help the reader to keep track of
the definitions, and to correlate at a glance the essential components of
the theory.

\section{Electric Stueckelberg Mechanism for Open $p$-Branes}
\label{opensec}

\subsection{Background}

In order to place our work in the right perspective, we begin this section
by briefly reviewing the way in which the interaction
of a $p$-brane with an antisymmetric tensor field can be described.
This interaction is nothing but a higher dimensional generalization of what is
usually done in the electromagnetic theory: there, we have a
$0$-dimensional object, namely a point particle, sweeping a
$1$-dimensional world--line. The most natural (i.e., ``geometric'') way
to couple a material particle to a field, is through its tangent element;
this is a vector field in the tangent bundle; it gives rise to a current
which, in turn, is coupled to a $1$-form potential. At this point, one may
contemplate a generalization of the electromagnetic scheme, and at least
two possibilities come to mind. Historically, the first extension of
electrodynamics was applied to non abelian gauge fields in order to include
internal symmetries with an eye on the weak and strong interactions, and
has paved the way to the formulation of the phenomenologically successful
theories embodied in the Standard Model of particle physics. With the
advent of string and membrane theory as the only paradigm capable, at least
in principle, of unifying gravity with the other fundamental interactions,
the possibility of a new generalization of the electromagnetic scheme has
emerged: in addition to the $p=0$ case, why not consider arbitrary values
of $p$ and have a theory of $p$-dimensional
objects endowed with $(p+1)$-dimensional tangent elements, interacting with
$(p+1)$-differential forms.
According to this universal blueprint, one might think that all {\it formal
properties} of electromagnetism
may be extrapolated to the extended case in a rather straightforward
manner. For {\it closed p--branes,} this is indeed the case\cite{acl} as
long as one keeps in mind that the physical properties of $p$--brane
electrodynamics, namely, the number of degrees of freedom of the object,
spin of the radiation field, etc., depend on the dimensionality $D$ of the
target spacetime, and, for fixed $D$, on the dimensionality of the
p--brane\cite{at},\cite{nip}. For instance, as mentioned before, in the
limiting case of ``bubble--dynamics'', $D=p+1$, there is no radiation
field, in contrast to the usual electrodynamics of point--charges. On the
other hand, in the case of {\it open p--branes} one may reasonably expect
some new formal, as well as physical properties that have no counterpart in
the electrodynamics of point charges. Evidently, such new properties derive
from boundary effects, regardless of the values of $D$ and $p$. The
universality of such boundary effects is of primary importance to us, since
they are deeply intertwined with the gauge invariance of p--brane
electrodynamics. In order to clarify that connection, let us call $J ^{\,
(p+1)}$ the current associated with the $p$-brane, and $A _{\, (p+1)}$
the corresponding $p+1$-tensor gauge potential.
If the $p$-brane is closed, that is, if its {\it world--manifold}, let us call
it ${\mathcal{E}}$, has no boundary, $\partial {\mathcal{E}} = \emptyset$,
then
the current $J$ is {\it divergenceless},
\beq
	\partial J ^{\, (p+1)} = 0
\eeq
and the corresponding action is invariant under the {\it tensor gauge
transformation}:
\beq
 	\delta _{\Lambda} A _{(p+1)}
    =
    d \Lambda _{(p)}
    \quad .
\label{gaugeuno}
\eeq
It seems worth emphasizing that ``charge conservation'' is now
\textit{associated with a topological property
of the extended object}, namely that it has no boundary. From here, one may
infer two things concerning the general case: from a mathematical
standpoint, one may expect that cohomology  plays a central role in the
theory, because one has to deal with differential forms of different order
defined on the bulk and boundary of the p--brane history; from a physical
standpoint, on the other hand, one may anticipate that the presence of a
boundary breaks gauge invariance thereby violating the above conservation
law. However, in the following subsection, we show that \textit{the
symmetry can be restored} by introducing \textit{a compensating field} of
the type originally suggested by Stueckelberg in the case of
point--particles\cite{stueck}.
This artifact brings into the theory a dimensionful coupling constant which
ultimately leads to a \textit{gauge invariant} generalized Higgs mechanism
for generating mass.\\

\subsection{Boundary Effect, Gauge invariance and Mass}

In this subsection, we go directly to the core of the problem, and discuss
in more detail the case of open $p$-branes endowed with electric charge
only, in order to illustrate in the simplest case the difference between the
open case and the closed one. As we have just seen, the source current of
an open $p$-brane is
not divergence--free because there is a leakage of current through the
boundary,
and this breaks the gauge symmetry of the action
\beq
    S [A , J]
    =
    \int
        \left [
            - \frac{1}{2}
            F ^{\, (p+2)} (A)
            F _{(p+2)} (A)
            +
            e _{p}\,
            A _{(p+1)}\,
            J ^{\, (p+1)}
        \right ]
    \quad .
\label{cloeleact}
\eeq
As a matter of fact, we now have
\beqa
	\delta S [A , J] _{A \to A + d \Lambda}
    & = &
    e _{p}
    \int
        d \Lambda _{(p)}
        J ^{\, (p+1)}
    \nonumber \\
	& = &
    e _{p}
    \int
        \Lambda _{(p)}\,
        \partial\, J ^{\, (p+1)}
    \nonumber \\
    & = &
    e _{p}
    \int
        \Lambda _{(p)}\,
        j ^{(p)}
    \neq 0
    \quad .
\eeqa
Furthermore, since the boundary of a $p$-brane is a world--manifold itself,
it can also be described in
terms of a current, say $j ^{(p),}$ which takes into account by how much
the source of the $p$-brane fails to be conserved. Mathematically, this means
$$
    j ^{(p)} = \partial J _{\mathrm{e}} ^{\, (p+1)}
    \quad .
$$
However, we wish to show that the gauge invariance of the theory can be
restored, along with the conservation law, by artificially ``closing'' the
boundary of the p--brane. The artifact that does the job, was discovered
long ago by Stueckelberg in his attempt to construct a gauge invariant
theory of a massive vector field\cite{stueck}. With hindsight, the
Goldstone boson which is instrumental for the working of the Higgs
mechanism, is nothing but a Stueckelberg compensating field that provides
the longitudinal degree of freedom necessary to turn a gauge field into a
massive vector field {\it without losing gauge invariance}. That old recipe
is the template that we wish to use, ``mutatis mutandis'' in the case of
open p--branes. Accordingly, we introduce a \textit{compensating
antisymmetric
tensor field} coupled to the boundary of the extended object by modifying
the  action (\ref{cloeleact}) as follows,
\beqa
	Z [ \, \overline{N} _{\mathrm{e}} \, ]
    & = &
    {1 \over Z [ \, 0 \, ]}
    \int [ \FID A ] [ \FID C ]
    	\,
		e ^{ - S [ \, A , C , J _{\mathrm{e}} \, ]}
	\nonumber \\
    S [ \, A , C , \overline{N} _{\mathrm{e}} \, ]
    & = &
    \int
		\left[
        	\,
            - { 1\over 2 }
            F _{(p+2)} ( A )\,
            F ^{\, (p+2)} ( A )
        \right.
		+
    \nonumber\\
	&	+ & e _{p}\left.
            \left(\,
            	A _{(p+1)}
                -
                d C _{(p)}\,
            \right)\,
			J _{\mathrm{e}} ^{\, (p+1)}
           +{ \kappa\over 2 }\left(\, A _{(p+1)}- d C _{(p)} \, \right) ^{2}
            \,
        \right]
    \quad .
	\label{open}
\eeqa
The above action is gauge invariant provided that the new field $C$ ( the
St\"uckelberg compensating field), responds to the transformation of the
gauge potential as follows,
\beqa
&&\delta A _{(p+1)}= d\Lambda _{(p)}\\
&&\delta C_{(p)}= \Lambda _{(p)}\label{newsym}  .
\eeqa
We wish to demonstrate, now, that this new invariant action, can be
rearranged to show how the additional interaction mediated by the
St\"uckelberg field, can be
traded off with a massive interaction among the $p$-brane elements.
 This can be seen as follows.\\
First, from the action (\ref{open}) we derive the field equations
    by means of variations with respect to the two gauge potentials, $A$
    and $C$, respectively:
    \beqa
        \partial\, F ^{\, (p+2)}
        +
        \kappa
        \left(
            A ^{\, (p+1)}
            -
            d C ^{(p)}
        \right)
        & = &
        e _{p}\,
        J ^{\, (p+1)}\label{maxw}
        \\
        \kappa
        \,
        \partial
        \left(
            A ^{\, (p+1)}
            -
            d C ^{(p)}
        \right)
        & = &
        e _{p}\,
        j ^{(p)}
\label{divfree}
        \quad .
    \eeqa
Second, we solve the two equations above
    in terms of currents and propagators. To this end, it is useful to
split $A$ and $J$
    into the sum of a divergenceless (hatted) part, and a curl--free
(tilded) piece:
    \beqa
    && J ^{\, (p+1)}= \hat{J} ^{\, (p+1)}+\widetilde{J} ^{\, (p+1)}\\
    && \partial\, \hat{J} ^{\, (p+1)}=0\ ,\quad d\,\widetilde{J} ^{\,
(p+1)}=0\\
    && \partial\, J ^{\, (p+1)}= j^ { (p) }\qquad\longrightarrow
       \widetilde{J} ^{\, (p+1)}=   d\,\frac{1}{\Box}\, j ^{(p)}\\
    && A ^{\, (p+1)}= \hat{A} ^{\, (p+1)}+\widetilde{A} ^{\, (p+1)}\\
    && \partial\, \hat{A} ^{\, (p+1)}=0\ ,\quad d\,\widetilde{A} ^{\,
(p+1)}=0\\
    && F _{(p+2)}=d\hat{A} _{(p+1)}\qquad\longrightarrow
    \hat{A} _{(p+1)}= \partial {1\over\Box } F _{(p+2)}
    \eeqa

    Then, from Eq.(\ref{divfree}) we obtain

    \beqa
    	\widetilde{A}^{\, (p+1)}- d C ^{(p)}
        & = &
        \frac{e _{p}}{\kappa} d\, { 1\over \Box }\, j ^{(p)}
    \eeqa

    while the Maxwell equation (\ref{maxw}) can be written in terms of
    the field strength as

    \beq
    \partial \left(\, { \Box + \kappa\over \Box }\, F ^{\, (p+2)}\, \right) =
    e _{p}\, \hat{J}^{\, (p+1)}\quad\longrightarrow
    F ^{\, (p+2)}= e _{p}\,  d\, { 1\over \Box +\kappa }\, \hat{J}^{\, (p+1)}
    \eeq

    Third, we substitute the solutions into the action (\ref{open})
    and obtain,
    \beqa
        S [\, J\, , j\,]
        &=& \int
            \left [ - { 1\over 2 } F _{(p+2)}   F ^{\, (p+2)} - {
\kappa\over 2
}
             F _{(p+2)}\, { 1 \over \Box } \,  F ^{\, (p+2)} - e _{p} \,
             F _{(p+2)}\, d{ 1 \over \Box }\hat{J}^{\, (p+1)}\,
            \right ]\nonumber\\
        &=& \int
            \left [\, -{ e _{p}^2\over 2 }
                \hat{J} ^{\, (p+1)}
                \frac{1}{\Box +\kappa }
                \hat{J} ^{\, (p+1)}
                + { e _{p}^2 \over 2\kappa }
                j ^{(p)}
                \frac{1}{\Box}
                j ^{(p)}
            \right ]
    \quad .
\label{sjj}
    \eeqa

Thanks to the extended gauge invariance (\ref{newsym}) introduced by the
Stueckelberg compensator, the action (\ref{sjj}) is written in terms
of {\it divergence--free currents.} The first term in the last line of
Eq.(\ref{sjj}) represents a
\textit{ short range, bulk interaction mediated by a massive field.}   On
the other hand, the last term in Eq.(\ref{sjj}) describes a {\it long range
  interaction confined to the boundary of the p--brane.} The above effects
represent the physical output of the Stueckelberg
 Mechanism for electric $p$--branes. To sum it up, one may trade the
presence of the boundary, when concentrating on the gauge invariance
properties of the theory, with an extra interaction mediated by the
compensator field, and reinterpret it as the propagation of massive degrees
of freedom
on the $p$-dimensional extended object. With hindsight, one also
  recognizes the long range interaction term in Eq.(\ref{sjj}) as the
residual trace of the interaction mediated
  by the massless Stueckelberg field $C_{(p)}$. This reminds us of a
``Meissner--type
  effect'' in which the compensator field is ``~expelled~'' form the bulk
and trapped on
  the boundary of the extended object. It has been suggested that
  in the limiting case $D=p+1$, that ``secret long range force'' can produce
  confinement in $D=2$, and glueball formation in $D=4$ \cite{lusch}.
  Accordingly, we expect similar effects in higher dimensions \cite{bags}.\\

In order to investigate the strong coupling
 dynamics  of this mechanism, we have to switch to the ``~magnetic phase~''
 of the model (\ref{cloeleact}). Indeed, from previous results in dual
models, one might expect that
 a strong electric coupling regime can be equivalently described in terms
 of a dual weak magnetic coupling phase.
Thus, in the next section, we turn our attention to the duality procedure
that will enable us to switch from the electric to the magnetic phase.

\section{Electric and Magnetic Branes}
\label{evsmsec}
\subsection{Extension of the Dirac Formalism}

In this Section we are interested in objects carrying electric as
well as magnetic charge. They are described by the following action,
whose origin can be traced back to the original work by Dirac on magnetic
monopoles\cite {dirac}:
\beq
	S [ \, A , G ; J \, ]
    =
    \int
    	\left [ \, - \frac{1}{2}
			\left(
            	F _{\, (p+2)} ( A )
            	-
				G _{\, (p+2)} (\, x\ ; n(\gamma)\, )
            \right) ^{2}
		+
        e _{p}\, A _{(p+1)} J ^{\, (p+1)}
		\,
        \right]
	\quad .
\label{actionem}
\eeq
Let us consider first the closed case as a testing ground for the subsequent
discussion of the open case. In the expression (\ref{actionem}) we have
explicitly separated the usual electromagnetic contribution to the field
strength, $F$, from the \textit{magnetic field strength} $G$, which is the
\textit{singular part} of the electromagnetic field due to the
presence of magnetic charge ($x= n(\gamma)$ represents the parent
$(D-p-3)$-Dirac--brane). The electric field strength
$F = d A$ originates as the exterior derivative of the electromagnetic
gauge potential $A$, which in turn is coupled to the
$(p + 1)$-dimensional history of the extended object (an electric $p$-brane)
through the source current $J$. Since this brane
represents a \textit{closed} object, its source
current $J$ is conserved and the full action is
gauge invariant under the transformation (\ref{gaugeuno}). Moreover, since
$d F = 0$, the Bianchi identities
are satisfied, and $F$ (as given by $A$) cannot describe magnetic sources.
If we insist in having magnetic charges in addition to the electric
ones, it is precisely those Bianchi identities that must be invalidated: in
fact, this is the role of the $G$ field, since the Bianchi identities that
follow from the action (\ref{actionem}) are not satisfied on
${\mathcal{M}}$, i.e., the
world--history of the magnetic brane.
Thus, to sum up the content of the action (\ref{actionem}),
we have a \textit{closed electric} $p$-brane source,
$J$, coupled to the tensor potential $A$ from which the $F$--
field originates, and a \textit{closed magnetic} $(D-p-4)$-Dirac--brane,
with source current $\overline{J}$, responsible for the violation of the
Bianchi identities for $F$. It may also be useful to say a few words about
the dimensionality of the various components:
the world manifold of a $p$-brane is $(p+1)$-dimensional, and its source
current is a
$(p+1)$-vector which couples to a $(p+1)$-form. Then the field is a
$(p+2)$-form whose Hodge dual is a $(D-p-2)$-form.
The divergence of the dual field equals the magnetic current,
which is thus a $(D-p-3)$-vector associated with the world--history
of a $(D-p-4)$-brane. Moreover, note that the magnetic brane, in Dirac's
formulation, is the boundary of a parent brane,
${\mathcal{N}}$, and thus it is necessarily
closed because the boundary of a manifold does not have a boundary.
The full action (\ref{actionem}) possesses  {\it two  distinct gauge
symmetries}: one is the original
gauge symmetry (\ref{gaugeuno}) which reflects the fact that the
\textit{electric} brane is closed, and under which $G$ is inert;
in addition, there is a new {\it magnetic gauge invariance}
under the combined transformations
\beqa
    G ^{\, (p+2)} ( x ; {\mathcal{U}} )
    & \longrightarrow &
	G ^{\, (p+2)} ( x ; {\mathcal{V}} )
	+
    g _{D-p-4 \, }\, d
    (\, {}^{\ast} \,\Omega\,) ^{\, (p+1)} ( x ; {\mathcal{O}} )
    \label{k1}
    \\
    A ^{\, (p+1)}
    & \longrightarrow &
    A ^{\, (p+1)}
    +
    g _{D-p-4 \, }
	\,
    (\, {}^{\ast} \Omega\, ) ^{\, (p+1)} ( x ; {\mathcal{O}} )
    \quad .
	\label{k2}
\eeqa
This new symmetry reflects the freedom in the choice of the \textit{parent
Dirac
brane}. More precisely, following Dirac's formulation, the freedom in the
choice of a parent brane can be interpreted in terms of the
gauge transformation given above
{\it provided that the electric and magnetic charges satisfy the Dirac
quantization condition}
\beq
 	e _{p}
    \,
    g _{D-p-4 \, }
    =
    2 \pi n
    \quad , \qquad
    (\hbox{in units $\hbar = c = 1$})
\label{dq}
\eeq
where, $n=1,2,3,\dots$. In the absence of the electric current, the
action (\ref{actionem}) is gauge invariant under (\ref{k1}) and (\ref{k2}).
However, in the presence of the $(A \, J)$-interaction term, the action
$S [ \, A , G ; J \, ]$ changes as follows
\beq
	\delta _{\Omega} S
    =
    (-1) ^{p(D-p)}
    \,
    e _{p}
    \,
    g _{D-p-4 \, }
    \,
    \int
		(\, {}^{\ast} \Omega\,) _{(p+1)} (x ; {\mathcal{O}})
        J ^{\, (p+1)} (\, x\ ; E \,)
	\quad .
\eeq
The target space integral of ${}^{\ast} \Omega J$ is an integer equal to the
number of times that the two branes intersect. Thus,
\beq
 	\delta _{\Omega} S
    =
    (-1) ^{p(D-p)}
    \,
    e _{p}
    \,
    g _{D-p-4 \, }
    \,
    \times
    \hbox{integer number}
    \quad .
\eeq
If the Dirac quantization condition holds, then
\beq
	\delta _{\Omega} S
    =
    \pm
    2 \pi
    \times
    \hbox{integer number}
    \quad .\label{dquant}
\eeq
Accordingly, the (Minkowskian) path--integral is unaffected by a
phase shift $S \rightarrow S \pm 2 \pi n$, and the gauge transformation
(\ref{k1}), (\ref{k2}), has no physical consequences on the
interacting system (\ref{actionem}).

\subsection{Duality Transformation for Open Electric and Magnetic Branes}
Our specific purpose, now, is to
extend the ``{\it{}dualization}'' procedure to a system of open
$p$-branes coupled to higher rank gauge potentials. The purpose of this
extension is to study how the dualization procedure, together with the
formalism developed above, affects the mass generation mechanism that we
have
proposed in Section \ref{opensec}.\\

We believe that the full impact of the dualization procedure cannot be
appreciated without keeping in mind some mathematical properties of the new
formalism that we are proposing in this paper. Thus, as a preliminary step,
we recall the results of the dualization procedure for closed objects. This
case is discussed in detail in subsection \ref{closedsec} of Appendix
\ref{dualityapp} using the path--integral method. The essential technical
point of that approach is that the partition function of a system
described, for instance, by the action
(\ref{actionem}) can be written in terms of the gauge invariant variable
$F$ in place
of the gauge potential $A$. To see this, we note that in the kinetic term
the gauge potential $A$ appears only through
its field strength  $F(A)$, while the interaction term can be written as a
constraint $N\cdot F(A)$, after an integration by parts, where $N\equiv
\partial J $
is an ``electric parent current'':

\beq
 	S _{\mathrm{INT}}
    =
    -
    e _{p}
    \int
    	\overline{N} ^{\, (p+2)}
        \,
        F_{(p+2)}( A)
    \quad .
	\label{Fintact}
\eeq

However, in switching to such a { \it field strength
formulation } one must exercise some extra care if extended magnetic
objects are
present. In that case, the field strength is the sum of the curl  $dA$ and
a ``singular magnetic field strength '' $G$  :
\beq
F(A)\longrightarrow \overline{F}=dA -G
\eeq
where $G$ is chosen in such a way that the magnetic brane current
$\overline{J}$
enters as a source
 in the ``Bianchi Identities'' for $\overline{F}$:
\beq
{ }^* d  \overline{F}= g \overline{J}  .
\label{newb}
\eeq

 The ``new'' Bianchi Identities (\ref{newb}) are encoded into the path integral
 through the following general relation:
\beq
\int [\,{\cal{D}} A\,]W[\, dA\ , J\ , \overline{J}\,]=
\int [\,{\cal{D}}\overline{ F}\,]  [\,{\cal{D}} n\,] [\,{\cal{D}}\overline{n}
\,]
\delta\left[\, { }^* d  \overline{F}- g\overline{J}\,\right]
\delta\left[\, \partial N - J \,\right]
\delta\left[\,   \partial \overline{N} - \overline{J} \,\right]
W[\, \overline{F}\ , N\ , \overline{J}\,]\label{master}
\eeq
where, $n(\xi)$ and $\overline{n}(\sigma)$ are the embedding functions of the
electric and magnetic parent branes respectively. Note that, for clarity,
we have suppressed all the
non essential labels in (\ref {master}). Indices, coupling constants and
numerical
factors have been reinserted in (\ref{new}).\\
The net result of the computations described in Appendix \ref{closedsec} is
the following expression for the dual action of (\ref{actionem}),
\beqa
	 S [ \, H , G , B , \overline{N} , \overline{J} \, ]
    =
     & - &
    \int
		\,
        \left[
        	\,
	        d B ^{\, (D-p-3)}
    	    -
        	\imath\,  (-1) ^{D(p+1)} e _{p}
	        \,
			(\, {}^{\ast} \overline{N}\, ) ^{\, (D-p-2)}
            \,
        \right] ^{2}
        +
     \nonumber \\
       \qquad
   & - &\imath\, \int\left[\, g _{D-p-4 \, }\, B _{(D-p-3)}\overline{J}
^{(D-p-3)}
     -e _{p}(\, {}^{\ast} N\, ) ^{\, (p+2)} \overline{N} _{(p+2)}\,\right]
    \quad .
\label{cdual}
\eeqa
The first term in (\ref{cdual}) is the  kinetic term for the dual field
strength. The regular part is the curl of the dual gauge potential, while the
singular part is the dual of the  electric brane current carrying the electric
brane on its boundary. This term is the dual of the original kinetic term in
(\ref{actionem}).\\
The second term in (\ref{cdual}) represents
the coupling between the dual potential and the {\it magnetic brane current}.
The
strength of the coupling is represented by the magnetic charge $g _{D-p-4
\, }$.
It is worth observing that the strength of this coupling is the inverse
of the original electric coupling thanks to the Dirac quantization condition
 (\ref{dquant}). Thus, an effect of the dualization procedure is to reverse
the value of the coupling constant. Hence, a system of strongly coupled
electric branes  can
be mapped into a system of weakly coupled magnetic branes and viceversa.
In current parlance, this is an example of ``{\it S--duality}'' connecting
the strong--weak coupling phases of a physical system \cite{dlect}.

Finally, the third term in (\ref{cdual}) describes a contact interaction
between
``{\it parent}'' branes. This term raises a potential problem: as we have
seen the
Dirac brane is a ``gauge artifact'' in the sense that its motion can be
compensated by an appropriate gauge transformation. However,
when the electric or magnetic brane coordinates are varied, extra
contributions may
enter the equation of motion leading to physical effects. In order to avoid
this inconsistency we have to invoke the ``Dirac veto'', namely, the condition
that the Dirac brane world surface should not intersect the world surface
of any
 other charged object. With this condition, no extra contribution comes to the
 equation of motion. In this connection, we must emphasize that the
 {\it physical object is the dual brane}, coupled to the $B$-potential, and not
  the Dirac brane which still maintains its pure gauge status.\\
Finally, note that the dual current $\overline{J} ^{\, (D-p-3)}$
is divergenceless. Hence, it has support over the world--manifold of a
{\it closed} $(D-p-4)$-brane and acts as a conserved source in the r.h.s.
of the $B$-field Maxwell equation
\beq
	\partial\left[\, d\, B^{\, (D-p-3)} - \imath\,  (-1) ^{p+1} e _{p}
	        \,
	(\, {}^{\ast} \overline{N}\, ) ^{\, (D-p-2)}  \,\right]
    = \imath\, g _{D-p-4 \, }\, \overline{J} ^{\, (D-p-3)}
    \quad .
\label{dualmaxw}
\eeq
In order to check the consistency of the above results, consider the two
systems defined by
equations (\ref{actionem}) and (\ref{cdual}) in a {\it vacuum}, i.e., when the
sources $J$ and $\overline{J}$ are switched off. In this case, we have two sets
of field equations and Bianchi identities. The first set is given by
\beqa
    \partial\, F ^{\, (p+2)} ( A )
    & = &
    0
    \quad , \qquad
	\hbox{Maxwell equations}
    \label{amaxw}
    \\
    d\, F ^{\, (p+2)} ( A )
    & = &
    0
    \quad , \qquad
	\hbox{Bianchi identities}
    \label{abianchi}
\eeqa
while the second set involves the $B$-field
\beqa
	& &
    \partial\, d\,  B^{\, (D-p-3)}
    =
    0
	\quad , \qquad
    \hbox{Maxwell equations}
    \label{bmaxw}
    \\
	& &
    d\, d\,  B^{\, (D-p-3)}
    =
    0
	\quad , \qquad
    \hbox{Bianchi identities}
    \quad .
    \label{bbianchi}
\eeqa
Thus, we recover the familiar result that the (classical) ``{\it duality
rotation}''
\beq
	F ^{\, (p+2)}
    =
	{} ^{\ast}
    d\,  B ^{\, (D-p-3)}
\eeq
exchanges the role of Maxwell equations and Bianchi identities:
\beqa
	\hbox{$A$-Maxwell equations}
    & \longleftrightarrow &
    \hbox{$B$-Bianchi identities}
    \\
	\hbox{$B$-Maxwell equations}
    & \longleftrightarrow &
    \hbox{$A$-Bianchi identities}
\eeqa
Of course, the same relations hold true if we switch on {\it both} the
electric and magnetic sources. In conclusion, by manipulating the
path--integral representation of
$Z [ \, J \, ]$ we have obtained the  duality relation between relativistic
extended objects of different dimensionality \cite{nepo}
\beq
	\widetilde{p}
    =
    D-p-4
    \quad ,
\label{chiuso}
\eeq
and the code of correspondence between dual quantities can be
summarized as follows
\beqa
 	A ^{\, (p+1)}
    & \longleftrightarrow&
    B ^{\, (D-p-3)}
    \\
	\hbox{Field equations (in vacuum)}
    & \longleftrightarrow &
	\hbox{Bianchi identities}
    \\
 	J ^{\, (p+1)}
    & \longleftrightarrow &
    (\overline{J}) ^{\, (D-p-3)}
    \\
	\hbox{closed\ } \hbox{$p$-brane}
    & \longleftrightarrow &
    \hbox{closed\ } \hbox{$(D-p-4)$-brane}
    \\
	Z [ \, J \, ]
    & \longleftrightarrow &
    Z [ \, I \, ]
    \quad .
\eeqa
For the convenience of the reader, a ``dictionary'' of the various
fields and currents in our model is listed in Table \ref{table2} and
Table \ref{table3}.

\begin{table}
\caption{Closed $p$-brane gauge potentials and fields.}
	\begin{tabular}{cccc}
		Field & Dimension in $\hbar=c=1$ units & Rank & Physical
Meaning
        \\
		\tableline
        \\
  		$A _{\, \mu _1 \dots \mu _{p+1}}$ & $(length) ^{1-D/2}$ &
$p+1$ & ``electric'' potential
        \\
  		$F _{\, \mu _{1} \dots \mu _{p+2}}$ & $(length) ^{-D/2}$ &
$p+2$
& ``electric'' field strength
        \\
  		$G _{\, \mu _{1} \dots \mu _{p+2}}$ & $(length) ^{-D/2}$ &
$p+2$
& ``magnetic'' field strength
		\\
  		$\overline{F} _{\, \mu _{1} \dots \mu _{p+2}}$ & $(length)
^{-D/2}$ &
$p+2$ &
  		``singular'' field strength
        \\
  		$B _{\, \mu _{1} \dots \mu _{D-p-3}}$ & $(length) ^{1-D/2}$ &
$D-p-3$ & dual gauge potential
		\\
  		$H _{\, \mu _{1} \dots \mu _{D-p-2}}$ & $(length)^{-D/2}$ &
$D-p-2$ & dual field strength
        \\
  		\\
	\end{tabular}
\label{table2}
\end{table}

\begin{table}
\caption{Closed $p$-brane associated currents}
	\begin{tabular}{cccc}
		Current Density & Dimension in $\hbar=c=1$ units & Rank &
Physical Meaning
        \\
		\tableline
        \\
		$J _{\, \mu _{1} \dots \mu _{p+1}}$ & $(length) ^{p+1-D}$ &
$p+1$& electric current
        \\
  		$\overline{N} _{\, \mu _{1} \dots \mu _{D-p-2}}$ & $(length)
^{-D/2}$
& $D-p-2$ & parent electric
current
        \\
		$\overline{J} _{\, \mu _{1} \dots \mu _{D-p-3}}$ & $(length)
^{-p-3}$
& $D-p-3$ & magnetic current
        \\
  		$N _{\, \mu _{1} \dots \mu _{D-p-2}}$ & $(length) ^{-D/2}$ &
$D-p-2$ & parent magnetic current
        \\
  		$\Omega _{\, \mu _{1} \dots \mu _{D-p-1}}$ & $(length)
^{1-D/2}$
& $D-p-1$ & gauge brane current
        \\
		\\
		\tableline
	\end{tabular}
\label{table3}
\end{table}

Finally, it seems worth observing that the dual action
(\ref{cdual}) is gauge invariant under ``magnetic gauge transformations''
\beq
	\delta _{\tilde{\Lambda}}
    B ^{\, (D-p-3)}
    =
    d \widetilde{\Lambda} ^{(D-p-4)}
    \quad .
 \eeq
Accordingly, $B$ is a {\it massless field} which is a solution of
the field equation (\ref{bmaxw}).

\section{Magnetic Stueckelberg Mechanism}

Given the formal apparatus outlined in the previous section, the discussion
of Section \ref{opensec} in which we applied the
Stueckelberg mechanism to restore gauge invariance in a theory of open
electric branes, can now be extended to
the case in which there is also a magnetic charge present. In the latter
case, we are dealing with the following objects:\\

\noindent
$(1)$ an electric open $p$-brane, with an electric closed $(p-1)$-boundary;\\
$(2)$ a magnetic open $(D-p-3)$-brane, with a magnetic closed
$(D-p-4)$-boundary.
\\

Therefore, the action for the full system is a natural generalization of
the action (\ref{open}):
\beqa
    S
    & = &
    \int
        \left[ -{1\over 2}\left(\, F _{\, (p+2)} (A)-
        G _{\, (p+2)}  \,\right)^2
        +
        e _{p}\,
        \left(\,
        	  A _{(p+1)}- d C _{(p)}
        \,\right)\,
        J ^{\, (p+1)} _{e}\,\right]
        +
    \nonumber \\
    & & \qquad +
    { \kappa\over 2 }\, \int\,
        \left(\, A ^{\, (p+1)} -  d C ^{(p)} \, \right) ^{2}\label{mact}
\eeqa
Once again, for the interested reader, the dualization procedure for this
action is described in detail in Appendix \ref{openapp}. Here, we merely
report the final result for the dual action,
\beqa
	Z [ \, \overline{N} _{\mathrm{e}} , \overline{J} _{\mathrm{g}} \, ]
    & = &
    {1 \over Z [ \, 0 , 0 \, ]}
	\int [ \FID \overline{D} ] [ \FID I ]
    	e ^{ - S [ \, \overline{D} ,\overline{J} _{\mathrm{g}} \, ]}
	\nonumber \\
	S [ \, D , B , \overline{N} _{\mathrm{e}} , \overline{J} _{\mathrm{g}}
	\, ] =
    & - &\frac{1}{2 \kappa} \int \left[\, d D ^{\, (D-p-2)}
            		-\imath\,  (-1) ^{D(p+1)-(p+1)^2} e _{p}
    	            \, {}^{\ast} \overline{N} _{e} ^{\, (D-p-1)}\,
    	            \right] ^{2}+\nonumber\\
    & + & {1\over 2}\int
        \left[\, D ^{\, (D-p-2)} + (-1) ^{Dp} d B ^{\, (D-p-3)}\,\right] ^{2}
	+\nonumber\\
	& - & \imath\,\int
            \left [\, (-1)^ {Dp}
                 N ^{ (D-p-2)}\, D _{(D-p-2)}
               +   g _{D-p-4 \, }\, B _{(D-p-3)}
                \overline{J} _{\mathrm{g}} ^{\, (D-p-3)}
                \,
            \right ]
    \quad .
\label{ampop}
\eeqa

Taking a closer look at the dual amplitude (\ref{ampop}), its most evident
feature is the {\it exchange of roles between compensator and gauge
field with respect to the electric phase:} presently, the
dual Stueckelberg strength tensor $D$ takes on the role of gauge
potential, while the dual gauge potential $B$ plays the role of
compensator. Together, these fields
implement the new gauge symmetry

\beqa
&&\delta_\Lambda D _{(D-p-2)}= d \Lambda_{(D-p-3)}\\
&&\delta_\Lambda B _{(D-p-3)}=  (-1) ^{Dp}\Lambda_{(D-p-3)}  .
\eeqa

Solving the field equations in the dual phase one arrives at the effective
action:
\beqa
	S[\,\widehat{ N}\ , \overline{J} _{\mathrm{g}}\ , J_e\,]=
	& - & {\kappa\over 2 }\int \widehat{N} ^{ (D-p-2)} { 1\over
	\Box +\kappa } \widehat{N} _{ (D-p-2)}+\nonumber\\
	&+ &\int \left[\, {  1\over 2 }\, g_{D-p-4 \, }^2
	\overline{J} _{\mathrm{g}} ^{\, (D-p-3)}
	{ 1\over \Box } \overline{J} _{\mathrm{g}\, (D-p-3)}
	+ {  e_p^2\over 2 \kappa} j _{\mathrm{e}} ^{\,(p)}
	{ 1\over \Box } j _{\mathrm{e}\,(p)}\,\right]
\eeqa

where the effect of the duality transformation becomes more transparent:
the bulk
interaction is short range while the boundary interaction is still long
range. On the basis of this result, we observe the following pattern of
duality relations

\beqa
&& A_{(p+1)}\longleftrightarrow D ^{\, (D-p-2)}\\
&& \widetilde{J}^{(p+1)} \longleftrightarrow \widehat{N} _{ (D-p-2)}\\
&& C_{(p)}\longleftrightarrow  B ^{\, (D-p-3)}\\
&&  {  e_p\over \sqrt \kappa} j _{\mathrm{e}} ^{\,(p)}\longleftrightarrow
 g_{D-p-4 }\overline{J} _{\mathrm{g}} ^{\, (D-p-3)}
\eeqa
In words, the above correspondence tells us that the gauge potential
$A_{(p+1)}$, which in the original phase is coupled to the electric
$p$-brane
current, transforms into $D ^{\, (D-p-2)}$ and interacts in a gauge invariant
way with a $D-p-3$-brane. The Stueckelberg mechanism induces a mass
$\sqrt\kappa$
for $A_{(p+1)}$, and the same mass for $D ^{\, (D-p-2)}$. In other words,
{\it both the electric and magnetic gauge potentials are massive due to the
mixing between different rank tensors.} This is the Stueckelberg mechanism
operating on higher order tensors rather than vectors and scalars.\\
In the original action (\ref{mact}), the gauge compensator is $C_{(p)}$, while
in the dual action the Stueckelberg field is the dual gauge potential
$B^{\, (D-p-3)}$. Overall, the dynamics of the model removes the compensator
field from the bulk spectrum, and confines it to the brane boundary where it
mediates a long-range interaction.\\
Finally, in the dual phase the electric parent brane disappears
as a {\it physical object.} Only  $\widehat{N} _{ (D-p-2)}$ and $\overline{J}
_{\mathrm{g}} ^{\, (D-p-3)}$ enter as  conserved sources in the
field equation. The only physical  electric source is the divergence free
current $j _{\mathrm{e}} ^{\,(p)}$ providing boundary electric/magnetic
duality symmetry. A similar phenomenon was discovered by Nambu to occur in the
Dual String Model of Mesons \cite{nambu}. Nambu showed that the mesonic open
string acquires physical reality when propagating in the non--trivial Higgs
vacuum because of the interaction between the end points  (magnetic monopoles)
and the charge of the scalar field. On our part, we have considered higher
dimensional open objects, and replaced the Higgs mechanism with the {\it
tensor mixing mechanism} as the basic engine of mass production at the
level of $p$-brane dynamics.

\section{Discussion of Results, Concluding Remarks and Outlook}
\label{concl}

A central issue to be confronted by any theory involving relativistic
extended objects is the search for an extension of the Higgs mechanism in
order to account not only for the mass spectrum of ordinary matter in the
form of elementary particles, but also for the existence of {\it cosmic
mass}, such as dark matter. A second central issue in the theory of
extended objects is to understand the role of ``duality'' in connecting
seemingly different realizations of the underlying $M$--theory. In this
paper we have addressed  both issues using the relatively familiar testing
ground provided by the ``electrodynamics of p--branes''. The method of
investigation that we have described in this paper, is an original
elaboration of the approach introduced in references\cite{seo}
and\cite{hk}, in order to study the strong coupling phase of the Higgs
model, and its relation with the dual string model. Moreover, the
derivation of Eq.(\ref{dq}) is an original extension of the method
introduced in Ref.\cite{klein} for pointlike charges and magnetic monopoles.\\
Our emphasis, throughout the paper, has been on the relationship between
the dimensionality of electric and magnetic p--branes with an eye on the
physical effects associated with the presence of a boundary for open
p--branes. A compendium of our results is encoded in the dual action
(\ref{ampop}) and in Table[\ref{table4}] where we have listed the relevant
properties of fields, currents and coupling constants for the open case. In
this
connection,
we observe that the current $K$ has support on the world manifold of a
$(D-p-2)$-brane. Hence, in contrast to the case of closed $p$-branes,
we conclude that the spatial dimensionality of dual open branes is given by
\beq
	\tilde{p} = D - p - 3  .
 \label{openp}
\eeq
It seems worth elaborating slightly on the physical meaning of this formal
relationship, since it reflects the central result of our discussion. Dual
open objects in $D=10$ spacetime are listed in Table[\ref{table5}].

\begin{table}
\caption{Open $p$-branes duality in $D=10$ spacetime dimensions.}
	\begin{tabular}{ccc}
	$p$ & $p$-brane &  $\tilde{p} = 7-p$ dual--brane
    \\
	\tableline
	\\
	$0$ & point--particle & $7$-brane
    \\
	$1$ & string & $6$-brane
    \\
 	$2$ & membrane & $5$-brane
    \\
	$3$ & bag & $4$-brane
    \\
	\\
	\end{tabular}
\label{table5}
\end{table}

That list is consistent with the results reported in Ref.\cite{suga}
through a different approach. By comparing Table \ref{table5} with
Table \ref{table1}, one sees at a glance that the pattern of dual objects in
$D = 10$ looks like the one for {\it closed} objects in $D = 11$ supergravity.
Far from being an accident, this similarity reflects the fact that a closed
$p$-dimensional surface can always be considered as
the boundary of an open $(p+1)$-dimensional volume. With hindsight,
it is not surprising that the relation (\ref{openp}) can be obtained from
(\ref{chiuso}) by replacing $p$ with $p+1$. However, the formal
relationship (\ref{openp}) reflects a deeper physical phenomenon, namely,
the generation of mass as a consequence of the presence of a physical
boundary. To be sure, the origin of a mass term is independent of the
dualization procedure. As we have shown in Section \ref{opensec}, it can be
traced back  to the introduction of the Stueckelberg field which is necessary
to compensate for the leakage of symmetry through the boundary of the
$p$-brane. A precursor of this mechanism was discussed, in the prehistory of
string theory, by Kalb and Ramond for an open string with a quark--antiquark
pair at the end points \cite{kalb}, and later extended to the case of an open
membrane having a closed string as its boundary \cite{aae}. Geometrically,
the introduction of the compensating field is tantamount to ``closing the
surface'', thereby restoring gauge invariance, albeit in an extended form. The
net result of the whole procedure is that the gauge field
acquires a mass through the ``mixing'' of different gauge potentials. While
the origin of mass is strictly a boundary effect, and therefore independent
of duality, it seems pertinent to ask how the mixing mechanism, i.e., the
relationship between mass, Stueckelberg field and gauge potential is
affected by the duality transformation. In order to answer this basic
question, already raised in the Introduction, in Appendix \ref{dualityapp} we
have shown how to implement the duality procedure in two distinct steps: the
duality transformation  is first applied  to the Stueckelberg sector of
the model, and then to the remaining gauge part of the  partially dualized
action. The final output is a massive, gauge invariant theory for higher
rank tensor fields and currents, written in terms of a dual gauge potential and
a dual Stueckelberg field. \\
The main result of that laborious work is a pair of (semi) classical effective
actions describing, in a gauge invariant way, the
interaction among electric and magnetic branes both in the original,
``electric'' phase, and in the dual, ``magnetic'' phase. In the two
actions, bulk and boundary dynamics are clearly separated. While the bulk
interaction is screened by the mass of the tensor gauge  field, the boundary
interaction is still long range. This is a somewhat unexpected result which
we interpret as a manifestation of the following {\it holographic principle}:
even when the Stueckelberg field is absorbed by the
gauge potential ( in analogy to the Goldstone boson in the Higgs model),
and therefore disappears as a physical excitation in the {\it bulk }of the
brane,
an imprint of the associated long range interaction is recorded on the
boundary as a reminder that {\it gauge invariance is rearranged, but not
lost.}\\
In conclusion, in view of the fact that the construction of a
Higgs Model for $p$-branes is at best tentative\cite{higgs}, with no
obvious way of spontaneously breaking gauge symmetry, at
present the only gauge invariant way to provide a
 tensor gauge field with a mass term is through the Stueckelberg mechanism
 discussed above. The {\it cosmological} implications of this new conversion
 mechanism that transforms the vacuum energy ``stored'' by a massless
 tensor gauge potential into massive particles, will be discussed in a
forthcoming publication\cite{stuck}. \\

\section{Note Added}
After this paper was accepted for publication we have been made
        aware of some related articles where p-brane electric/magentic duality 
        and Stueckelberg/Higgs mechanism was addressed in a similar form
        \cite{last}

\appendix
\section{Interacting, closed electric and magnetic branes}
\label{dualityapp}

\subsection{Closed Branes}
\label{closedsec}

The precise meaning of the term ``duality'' is implicitly assigned by the
procedure employed in this subsection. It can be summarized thus:\\

\noindent
$(1)$ to exchange the gauge potential $A$ in favor of the field strength
$F$, by introducing the Bianchi identities as a constraint in the
path--integral
 measure;\\
$(2)$  to introduce a ``dual'' gauge potential $B$ as the ``Fourier
    conjugate'' field to the Bianchi identities;\\
$(3)$  to integrate out $F$, and identify the dual current as the
	object linearly coupled to $B$.\\

Step(1). In order to write the current--potential interaction in
terms of $F$, we note that, since $J$ has vanishing divergence, it is
a {\it boundary current} and can be written as the divergence of a
``{\it parent}'' electric current. In other words, there exists a
$(p+2)$-rank current
$\overline{N} ^{\, \mu _{1} \dots \mu _{p+2}} \left( x\ ; \overline{n}\right)$
such that
\beqa
 & &   \overline{N} ^{\, \mu _{1} \dots \mu _{p+2}} ( x \ ; \overline{n} )
     =
	\int _{\overline{\Gamma}} d ^{p+2} \overline{\gamma}\,
     	d\overline{n} ^{\mu _1}\wedge \dots \wedge d\overline{n} ^{\mu _{p+2}}
        \,\delta ^{D)}\left[\, x -
        \overline{n} \left( \overline{\gamma} \right)\,\right]
    \\
 &  &	\partial _{\, \mu _{1}}
    \overline{N} ^{\, \mu _{1} \mu _{2} \dots \mu _{p+2}}
     =
	J ^{\, \mu _{2} \dots \mu _{p+2}}
    \quad .
	\label{parent}
\eeqa
To understand the role of these currents, it is useful to recall once
again Dirac's construction in which a particle anti--particle pair can be
interpreted as the boundary of an ``electric string'' connecting
them \cite{dirac}.  In our case, which involves higher dimensions, the
analogue of a particle anti--particle pair is a closed
$p$-brane which we interpret as the boundary of an {\it open},
$(p+1)$-dimensional, parent brane. From this vantage point,
the interaction term  in the action can be written as follows
\beq
 	S _{\mathrm{INT}}
    =
    -
    {e _{p} \over (p+2) \, !}
    \int
    	\overline{N} ^{\, (p+2)}
        \,
        F _{(p+2)}(A)
    \quad .
\eeq
Incidentally, note that one may employ the same procedure
 directly in the expression for the ``{\it{}non}--magnetic''
action (\ref{actionem}), and write it only in terms of the field strength
$F$ as follows
\beqa
	&& S [\, A\  , G\ , \overline{N} \, ]
    =
    \int
    \left[
    	\,
        - \frac{1}{2}
        \overline{F} ^{\, (p+2)}
        \overline{ F} _{(p+2)} +
        e _{p}\,
    	\overline{N} ^{\, (p+2)}\, \left(\,  \overline{F} _{(p+2)} +
    	   G_{(p+2)}\,\right) \, \right]
    \quad .
    \label{fieldaction}\\
&& \overline{F} _{(p+2)} ( A )\equiv F _{(p+2)} ( A )- G_{(p+2)}
\eeqa
Proceeding with the dualization procedure, we further note that since the
interaction term in the action (\ref{actionem}) is written
in the form (\ref{Fintact},) one may also use $\overline{F}$ as integration
variable in the functional integral. One can do that, i.e., treat
$\overline{F}$
as an {\it independent variable}, by imposing the Bianchi identities as a
constraint. The relationship between $\overline{N}$ and $J$ must also be
encoded in the path--integral . This can be achieved by performing an
integration over the parent brane coordinates
$\overline{N} = \overline{N} ^{\mu} \left( \overline{\gamma} ^{i} \right)$,
which are constrained to satisfy equation (\ref{parent}). These steps are
implemented by inserting the following (functional) equivalence relation into
the path--integral
\beqa
  	\int [ \FID A ] & &
    	\,W [ \, dA\ , G\ , J\ ,  \overline{J}\, ]
     =
  	\int [ \FID \overline{F} ] [ \FID n ] [ \FID \overline{n} ]
  	\delta\left[\, d\, {}^* \overline{F} _{\, (p+2)} - g _{D-p-4 \, }
         \overline{J} ^{\, (D-p-3)}
    \right]
    \times
    \nonumber\\   \times & &
    \delta\left[\,
        \partial N ^{\, (D-p-2)} - g _{D-p-4 \, }\overline{J} ^{\, (D-p-3)}
        \,\right]\delta
    \left[\, \partial \overline{N} ^{\, (p+2)} - J ^{\, (p+2)} \,
    \right]\, W[\, \overline{F}\ , N \ , \overline{N} \, ]
	\quad .
 	\label{new}
\eeqa
The first Dirac delta--distribution takes into account the presence of the
magnetic brane as the singular surface where the Bianchi identities are
violated. The second and third delta functions encode the relationship between
the boundary currents $J$, $\overline{J}$ and their  respective  bulk
counterparts $\overline{N}$ and $N$. It may be worth emphasizing that the
electric and magnetic parent branes enter the path--integral as  ``dummy
variables'' to be summed over. In other words, the {\it physical sources} are
the boundary currents $J$ and $\overline{J}$ alone.
After performing the above operations, the generating functional, written
in terms of the total electric--magnetic field strength $\overline{F}$, takes
the form
\beqa
    Z [ \, J , \overline{J} \, ]
    &=&
    {1 \over Z [ \, 0 \, ]}
    \int [ \FID \overline{F} ] [ \FID N ] [ \FID \overline{N} ]
    	\,
	\delta\left[\,{}^{\ast}d \overline{F} ^{\, (p+2)}
            - \overline{J} ^{\, (D-p-3)}
            \,\right]
        \times
	\nonumber \\
	& & \qquad \times
        \delta
        \left[\,
            \partial \overline{N} ^{\, (p+2)}- J ^{\, (p+2)}\,\right]
		\delta
        \left[
        	\,
            \partial N ^{\, (D-p-2)}
            -
            g _{D-p-4 \, } \overline{J} ^{\, (D-p-2)}
            \,
        \right]
        \,
        e ^{ - S [ \, \overline{F} , G , \overline{N} \, ]}
	\label{vacuumf}
        \eeqa
Step(2). The ``Bianchi identities Dirac delta--distribution''  can be Fourier
transformed  by means of the functional representation
\beq
	\delta
    \left(\, {}^*
        d \overline{F} ^{\, (p+2)}
        -
  		g _{D-p-4 \, } \, \overline{J} ^{\, (D-p-3)}
    \right)
    =
	\int [ \FID B ] \,
    	\exp
        \left(
        	\imath\,
        	\int
            	\, L ( B , \overline{F} , \overline J)
        \right)
    \quad ,
    \label{fourier}
\eeq
where
\beq
	L ( \, B , \overline{F} , \overline{J} \, )
    =  B _{\, (D-p-3)}\, \left[\,
       {}^{\ast} d \overline{F} ^{\, (p+2)}- \, g _{D-p-4 \, }
         \overline{J} ^{(p+3)}\,\right]   \quad .
\eeq
The net result is a ``field strength formulation''  of the model
(\ref{vacuumf}, \ref{actionem}) in terms of a dynamical field $F$, and
a Lagrange multiplier $B$:
\beqa
	Z [ \, J , \overline{J} \, ]
    & = &
    {1 \over Z [ \, 0 \, ]}
    \int [ \FID N ] [ \FID \overline{N} ] [ \FID \overline{F} ] [ \FID B ]
    	\delta
        \left [
        	\partial N ^{\, (D-p-2)}
            -
            g _{D-p-4 \, } \overline{J} ^{\, (D-p-3)}
        \right ]
        \times
    \nonumber \\
    & & \qquad \qquad \times
		\delta
        \left [
        	\partial \overline{N} ^{\, (p+2)}
            -
            J ^{\, (p+2)}
        \right ]
        \,
        e ^ { - S [ \, \overline{F} , G , B , \overline{N} \, ]}
    \quad ,
\label{primord}
\eeqa
where
\beqa
S [ \, \overline{F}\ , G\ , B\ , \overline{N}\ , \overline{J} \, ]
    =& - &\int\left[\,
		{ 1\over 2}\overline{F} ^{\, (p+2)} \overline{F} _{(p+2)} +
            e _{p}\, \left(\overline{F} _{(p+2)}+ G _{(p+2)}
            \right)
			\overline{N} ^{\, (p+2)}
			+
        \right .
    \nonumber\\
	  	& - & \imath\, (-1)^ { D(p+1) }\left .
	    	{}^* H ^{\, (p+2)}(B)\,\overline{F} _{(p+2)}
        		-\imath\,   g _{D-p-4 \, } \,
		     B _{(D-p-3)}\, \overline{J} ^{(D-p-3)}
	    	\,
		\right]
\eeqa
and ${}^{\ast} H$ represents the {\it dual field strength}
\beq
	 H _{\, (D-p-2)}(B)\equiv d B _{\, (D-p-3)}
    \quad .
\eeq
Step(3).
Finally, we are ready  to switch to the dual description of the model by
integrating away the field strength $\overline{F}$.
Since the path--integral is Gaussian in $\overline{F}$,
the integration can be carried out in a closed form:
\beq
 Z [ \, J , \overline{J} \, ]
     = {1 \over Z[ \, 0 \, ]}
    \int [ \FID N ] [ \FID \overline{N} ] [ \FID B ]
    	\,
        \delta \left[ \partial N - g _{D-p-4 \, } \overline{J} \right]
        \,
		\delta
        \left[
            \partial
            \overline{N}
            -
            J
        \right]
        \,
        e ^{ - S [ \,  G\ , B\ , \overline{N}\ , \overline{J} \, ]}
        \eeq
	\beqa
	S [ \,  G\ , B\ , \overline{N}\ , \overline{J} \, ]
    =   &-&  {1\over 2} \int\, \left[ \, i(-1)^ { D(p+1)}
       (\, {}^{\ast} H\,) ^{\, (p+2)}+ e _{p}\, \overline{N} ^{\, (p+2)}\,
        \right] ^{2}+\nonumber\\
       &  - & \int\, \left[\, e _{p}\,
            G _{(p+2)}
            \overline{N} ^{\, (p+2)}
            + i\, g _{D-p-4 \, }\, B _{(D-p-3)}\, \overline{J} ^{(D-p-3)}
            \,\right]
    \nonumber\\
     =
    & - &
    {1\over 2}\int
		\,
        \left[
        	\,
	        d B _{\, (D-p-3)}
    	    -
        	\imath\,  (-1) ^{D(p+1)} e _{p}
	        \,
			{}^{\ast}( \overline{N} ^{\, (p+2)})
            \,
        \right] ^{2}+ \nonumber \\
        & + & \imath\, \int \left[\,  g _{D-p-4 \, }
             B _{(p+3)}\, \overline{J} ^{(p+3)}
           +    e _{p}(\, {}^{\ast} N) ^{\, (p+2)}
            \overline{N} _{(p+2)}\, \right]
    \quad .
\eeqa

\subsection{Open Branes}
\label{openapp}

Equipped with the formalism developed in the previous subsection,
we now wish to consider the extension
of the path--integral method to the case of open $p$-branes. As emphasized
throughout the paper, the main difference stems from the fact that a gauge
invariant
action for an open $p$-brane requires the introduction of new gauge fields to
compensate for  the gauge symmetry ``leakage'' through the boundary
\cite{at}. Thus, we replace the system (\ref{cloeleact})
with the following one,
\beqa
	Z [ \, \overline{N} _{\mathrm{e}} \, ]
    & = &
    {1 \over Z [ \, 0 \, ]}
    \int [ \FID A ] [ \FID C ]
    	\,
		e ^{ - S [ \, A , C , J _{\mathrm{e}} \, ]}
	\nonumber \\
    S [ \, A , C , \overline{N} _{\mathrm{e}} \, ]
    & = &
    \int
		\left[\,  -\frac{1}{2}
           \overline{F} ^{\, (p+2)}\, \overline{F}_{(p+2)} + e _{p}\,
            \left(\, A _{(p+1)} - d C _{(p)}\,\right)
	\overline{N} _{\mathrm{e}} ^{\, (p+1)}
        \right.
		+
    \nonumber\\
	& & \qquad \qquad \qquad\qquad
    	\left.
        	-
            \kappa
			\left(
                d C _{(p)}
                -
                A _{\, (p+1)}
            \right) ^{2}
            \,
        \right]
    \quad .
\eeqa

where $C^ { (p) } $ has canonical dimensions $L ^{2-D/2}$,
and $\kappa$ is a constant introduced for dimensional reasons.

In this case, the divergence of the $p$-brane current is no longer
vanishing, but equals the current $J _{\mathrm{e}} ^{(p)}$
associated with the free boundary of the world--manifold. In other words,
\beq
	\partial \overline{N} _{\mathrm{e}} ^{\, (p+1)}
    =
	J _{\mathrm{e}} ^{(p)}
    \quad .
\eeq

However, the action (\ref{open}) is still invariant under the
{\it extended gauge transformation}:
\beqa
	\delta _{\Lambda} A _{\, (p+1)}
    & = &
    d \Lambda _{(p)}
    \\
	\delta _{\Lambda} C _{(p)}
    & = &
	\Lambda _{(p)}
    \quad .
\eeqa

Indeed, the role of the $C_{(p)}$-field, which is a St\"uckelberg
compensating field, is to restore
the gauge invariance broken by the boundary of the $p$-brane.
Perhaps it's worth emphasizing that $S[\, A , C , \overline{N}
_{\mathrm{e}} \,]$ depends on $C_{(p)}$ only through its covariant curl
$d C  _{(p)}\equiv \Theta_{(p+1)} $, which {\it is not gauge invariant}, but
transforms as follows
\beq
	\delta _{\Lambda} \Theta_{(p+1)}
    =
    d \Lambda _{(p)}
    \quad .
\eeq
The advantage of the path--integral method for constructing the dual
action becomes evident at this point, since we can eliminate the
Stueckelberg  potential $C _{(p)}$ in favor of its curl
$\Theta _{\, (p+1)}$ by introducing the {\it dual Stueckelberg potential }
$D^{(D-p-2)}(x) $  as we did in equation (\ref{fourier})
\beqa
    \delta
    \left[\, {}^* d\, \Theta _{\, (p+1)}\, \right]
    & = &
	\int [ \FID D ]
    	e ^ {\imath\, S [ \, D\ , \Theta \, ] }
    \\
    S [ \, D \ , \Theta \, ]
    & = &\int\,  D ^{\, (p+2)}
		\left(\, {}^{\ast}d\,\Theta _{\, (p+1)}\,\right)
         \nonumber\\
	& = & (-1)^ { Dp }
    \int \, \Theta ^{\, (p+1)} \left(\, {}^{\ast} K(D)\,\right) _{(p+1)}\\
    \left(\, {}^{\ast} K(D)\,\right) _{(p+1)} & = &
     \left(\, {}^{\ast} d\, D _{D-p-2}\,\right)
   \quad .
\eeqa

The resulting vacuum amplitude is
\beq
	Z [ \, \overline{N} _{\mathrm{e}} \, ]
    =
    {1 \over Z [ \, 0 \, ]}
    \int [ \FID A ] [ \FID D ] [ \FID I ]
    	e ^{ - S [ \, A \ , D \ , \Theta , \overline{N} _{\mathrm{e}} \, ]}
    \quad ,
\eeq
where
\beqa
	S [ \, A\ , D\ , \Theta\ , \overline{N} _{\mathrm{e}} \, ]
    &=&
    \int\left[\,
            - { 1\over 2 }\overline{F} _{(p+2)} \,\overline{ F} ^{\, (p+2)}
        	+
            e _{p}\, \left(\, A _{(p+1)}- \Theta _{\, (p+1)}\, \right)
			\overline{N} _{\mathrm{e}} ^{\, (p+1)}
        \right.
		+
    \nonumber\\
	 \qquad
    	   & + &
	       { \kappa\over 2 }\left.
			\left(
    	    	\Theta _{\, (p+1)}- A _{\, (p+1)}
    	    \right) ^{2}
			+
	        \imath\,  ( - 1 ) ^{Dp}
        	\Theta ^{\, (p+1)}
    	    \left(\, {}^{\ast} K( D )\,\right) _{\, (p+1)}
	        \,
        \right]
\eeqa

represents the ``{\it Stueckelberg} Dual Action'', in the sense that the
dualization procedure was applied to the field $C_ { (p)} $ only.\\

Translational invariance of the functional integration measure
enables us to shift the $I$-field and introduce the
{\it gauge invariant field strength} $\overline{\Theta}$
\beq
	\overline{\Theta} ^{\, (p+1)}
    \equiv
     \Theta^{\, (p+1)}
	-
    A ^{\, (p+1)}
\eeq
as a new integration variable instead of  $\Theta$.\\

Once $\overline{\Theta}$ is integrated out, we find

	\beq
	Z[\, \overline{N} _{\mathrm{e}} \, ]
	=
	{1 \over Z[ \, 0 \, ]}
	\int [ \FID A ] [ \FID D ]
		e^{-S[\, A\ , D\ , \overline{N} _{\mathrm{e}} \, ]}
	\eeq

with

	\beqa
	S [ \, A\ , D\ , \overline{N} _{\mathrm{e}} \, ]
    & = &
    \int\left[\, -\frac{1}{2} \overline{F} ^{\, (p+2)}\overline{F} _{(p+2)}
		+\imath\, (-1)^ { p(D-p) }
        (\, {}^{\ast} D\, ) ^{\, (p+2)}\left(\,
       \overline{F} _{(p+2)}+ G _{(p+2)}\,\right)\right.
	+
 	\nonumber\\
	& & \quad \qquad
    \left.
    	+
        \frac{1}{2\kappa}\left(\, e _{p} \,
            \overline{N} _{\mathrm{e}} ^{\, (p+1)} -
            \imath\,  ( -1 ) ^{Dp}(\, {}^{\ast} K) ^{\, (p+1)}\, \right)^2 \,
    \right]
    \quad .
	\label{mha}
\eeqa

By recognizing that the first line in equation (\ref{mha}) is the
same as (\ref{fieldaction}), once $ {}^{\ast} D$
is identified with $\overline{N}$, i.e.,
\beq
	e _{p}\, \overline{N} ^{\, (p+2)}_{\mathrm{e}}
        \rightarrow
	\imath\,  (-1) ^{(D-p)p} \, {}^*(\,  D _{\, (D-p-2)}\, )
    \quad ,
\eeq
we can write the  ``{\it Complete} Dual Action''
 without repeating all the previous calculations:
\beqa
	Z [ \, \overline{N} _{\mathrm{e}} , J _{\mathrm{g}} \, ]
    & = &
    \frac{1}{Z [ \, 0 , 0 \, ]}
	\int [ \FID B ] [ \FID D ] [ \FID w ]
    	\,
		\delta
        \left[
        	\,
            \partial \overline{N}^ { D-p-2 } -g \, \overline{J}^ { D-p-3 }
            \,
        \right]
 		e ^{ - S [ \, B , \, D , \overline{N} _{\mathrm{e}} \, ]}
	\nonumber \\
  S [ \, G , B , \overline{N} _{\mathrm{e}} , \overline{J} _{\mathrm{g}} \, ]
   &  = &
    \frac{1}{2 \kappa} \int \left[\, d D ^{\, (D-p-2)}
            		+\imath\,  (-1) ^{D- (p+1)^2} e _{p}
    	            \,(\, {}^{\ast} \overline{N} _{e}) ^{\, (D-p-1)}\,
    	            \right] ^{2}+\nonumber\\
    & - & {1\over 2}\int
        \left[\, D ^{\, (D-p-2)} -  (-1) ^{Dp} d B ^{\, (D-p-3)}\,\right] ^2 +
	\nonumber\\
	& - &
    	\imath\int
          \left [\, (-1)^{Dp} N ^{ (D-p-2)}\, D _{(D-p-2)}
               +  g _{D-p-4 \, }\, B _{(D-p-3)}
                \overline{J} _{\mathrm{g}} ^{\, (D-p-3)}\,
            \right ] \quad .\label{chefatica}\\
\eeqa

>From the dual  action (\ref{chefatica}) one gets the field equations

\beqa
&& \partial\, \left[ \, D ^{(D-p-2)} - (-1)^{Dp} d B ^{(D-p-1)}\,\right]=
-\imath (-1)^ { Dp } g_{D-p-4 \, } \overline{J} _{\mathrm{g}} ^{\, (D-p-3)}
\label{dual1}\\
&& \partial\, \left[ \,d D ^{\, (D-p-2)}
            		+\imath\,  (-1) ^{D- (p+1)^2} e _{p}
    	            \,(\, {}^{\ast} \overline{N} _{e}) ^{\, (D-p-1)}\,
    	 \right] + \kappa D ^{\, (D-p-2)}= -\imath\,\kappa (-1)^{Dp}
    	 N ^{ (D-p-2)}
\label{dual2}
\eeqa

Solving equation (\ref{dual1}), one finds
\beqa
&& D ^{(D-p-2)} - (-1)^{Dp} d B ^{(D-p-1)}=
\widehat{D} ^{(D-p-2)} -\imath (-1)^ { Dp } g_{D-p-4 \, } d {1\over\Box }
\overline{J}_{\mathrm{g}} ^{\, (D-p-3)}\\
&& \partial\,\widehat{D} ^{(D-p-2)}=0\label{BgaugeD}
\quad .
\eeqa

After decomposing the magnetic parent current as follows

 \beq
 N ^{ (D-p-2)}= \widehat{N} ^{ (D-p-2)} +  g_{D-p-4 \, } d {1\over\Box }
 \overline{J} _{\mathrm{g}} ^{\, (D-p-3)}\ ,\qquad \partial\,\widehat{N} ^{
 (D-p-2)}=0
 \eeq

equation (\ref{dual2}) becomes

 \beq
 \partial\, d\, \widehat{D} ^{\, (D-p-2)} + \kappa \widehat{D} ^{\, (D-p-2)}=
 \imath\, \kappa (-1)^{Dp} \widehat{N} ^{ (D-p-2)}\label{sol1}
 \eeq

and gives for $\widehat{D} ^{\, (D-p-2)}$ the following solution

   \beqa
   &&\widehat{D} ^{\, (D-p-2)}= -\imath\,\kappa (-1)^{Dp}
    {1\over \Box +\kappa}\widehat{N} ^{ (D-p-2)}\label{sol2}\\
 &&  d\, \widehat{D} ^{\, (D-p-2)}=-\imath\,\kappa (-1)^{Dp}
        d {1\over \Box +\kappa}\widehat{N} ^{ (D-p-2)}
\quad .
\label{sol2bis}
	\eeqa

When (\ref{sol1}), (\ref{sol2}) are inserted back into (\ref{chefatica})
we find the following expression
\beqa
	S[\,\widehat{ N}\ , \overline{J} _{\mathrm{g}}\ , J_e\,]=
	& - & {\kappa\over 2 }\int \widehat{N} ^{ (D-p-2)} { 1\over
	\Box +\kappa } \widehat{N} _{ (D-p-2)}+\nonumber\\
	&+ &\int \left[\, {  g_{D-p-4 \, }^2\over 2 }
	\overline{J} _{\mathrm{g}} ^{\, (D-p-3)}
	{ 1\over \Box } \overline{J} _{\mathrm{g}\, (D-p-3)}
	+ {  e_p^2\over 2 \kappa} j _{\mathrm{e}} ^{\,(p)}
	{ 1\over \Box } j _{\mathrm{e}\,(p)}\,\right]
\eeqa

in which the short-range bulk interaction and the long range boundary
interaction are clearly
displayed.
\begin{table}
\caption{Open $p$-brane gauge potentials and fields.}
	\begin{tabular}{cccc}
		Field & Dimension in $\hbar=c=1$ units & Rank & Physical
Meaning
        \\
		\tableline
        \\
		$A _{\, \mu _{1} \dots \mu _{p+1}}$ & $(length) ^{1-D/2}$ &
$p+1$ & ``electric'' potential
        \\
		$F _{\, \mu _{1} \dots \mu _{p+2}}$ & $(length) ^{-D/2}$ &
$p+2$
& ``electric'' field strength
        \\
		$C _{\, \mu _{1} \dots \mu _{p}}$ & $(length) ^{2-D/2}$ &
$p$ & St\"uckelberg gauge potential
        \\
		$K _{\, \mu _{1} \dots \mu _{p+1}}$ & $(length) ^{1-D/2}$ &
$p+1$ & St\"uckelberg field strength
		\\
		$D _{\, \mu_{p+3}\dots \mu_D}$ & $(length) ^{-D/2}$ &
$D-p-2$ &
dual St\"uckelberg  potential
        \\
		$I _{\, \mu_1\dots\mu_{p+1}}$  & $(length) ^{-1-D/2}$ &
$p+1$ &
dual St\"uckelberg strength\\
\\
	\end{tabular}
\label{table4}
\end{table}

\begin{table}
\caption{Open $p$-brane associated currents}
	\begin{tabular}{cccc}
		Current Density & Dimension in $\hbar=c=1$ units & Rank &
Physical Meaning
        \\
		\tableline
       \\
       $\overline{N} _{\mathrm{e}}^{ \, \mu _{1} \dots \mu _{D-p-2}}$ &
$(length) ^{p+1-D}$ & $p+1$ &
electric parent current
        \\
	$J _{\mathrm{e}} ^{\, \mu _{1} \dots \mu _{p}}$ & $(length)^{p-D}$
	& $p$& boundary current
        \\
  	$N ^{\, \mu _{1} \dots \mu _{D-p-2}}$ & $(length) ^{-D/2}$ &
$D-p-2$ &  parent magnetic current
        \\
  	$\overline{J} _{\mathrm{g}}^{ \, \mu _{1} \dots \mu _{D-p-3}}$ &
$(length) ^{-p-3}$ & $D-p-3$ &
magnetic current	\\
\\
		\tableline
	\end{tabular}
\label{table6}
\end{table}

\begin{table}
\caption{Dimension of couplings}
	\begin{tabular}{cccc}
		Coupling constants & Dimension in $\hbar=c=1$ units
		\\
		\tableline
        \\
		$e$ & $(length) ^{D/2-p-2} $
        \\
		$g$ & $(length) ^{p+2-D/2}$
        \\

		$e \, g$ & $1$
        \\
		$\kappa$ & $(length) ^{-2}$	\\
	\\
	\tableline
	\end{tabular}
\label{couptable}
\end{table}


\begin{thebibliography}{99}
\bibitem{nepo}
    	R.Nepomechie,
    		Phys.Rev. {\bf D31}, 1921, (1984);\\
		C.Teitelboim,
        	Phys.Lett. {\bf B167}, 69, (1986)
\bibitem{bags}
	S.Ansoldi, A.Aurilia, A.Smailagic, E.Spallucci,
	``~Dualization of Interacting Theories Including $p=d-1$ Limiting
	Cases,~'' hep-th/9911051,  Physics Letters B, in press
\bibitem{acl}A.Aurilia, D.Christodoulou, F.Legovini,
	Phys.Lett. {\bf 73B,} 429, (1978)
\bibitem{at}A.Aurilia, Y.Takahashi,
        	Progr.Th.Phys. {\bf 66,} n.2, 693, (1981)
\bibitem{nip}Z. Tokuoka, Phys. Lett. {\bf 87A,} 215,(1982)
\bibitem{stueck}E.C.G. Stueckelberg, Helvetica Physica Acta {\bf 11}, 225,
(1938)
\bibitem{lusch} M.L\"usher,
	Phys. Lett. {\bf 78B,} n.4, 465, (1978)
\bibitem{dirac}
    	P.A.M. Dirac,
			Phys.Rev. {\bf 74}, 817 (1948)

\bibitem{dlect}S.E. Hjelmeland, U. Lindstr\"{o}m,
	``Duality for the Non--Specialist'' hep-th/9705122
\bibitem{nambu}
	         Y.Nambu,
	 Phys.Rev. {\bf D10}, 4262, (1974)
\bibitem{seo}
    	A.Sugamoto,
        	Phys.Rev. {\bf D19}, 1820, (1979);\\
		K.Seo, M.Okawa, A.Sugamoto,
        	Phys.Rev. {\bf D19}, 3744, (1979)
\bibitem{hk}
    	H.Kawai,
        	Progr.Theor.Phys. {\bf 65}, No.1, 351, (1981)
\bibitem{klein}
	H.Kleinert,
	Int. J. of Mod. Phys. {\bf A7,} No.19 (1992) 4693
\bibitem{suga}
 		A.Sugamoto,
        	``Old--fashioned dualities revisited'', hep-th/96111051
\bibitem{kalb}
    	M.Kalb, P.Ramond,
			Phys.Rev. {\bf D9}, 2273, (1979)

\bibitem{aae}A.Aurilia, E.Spallucci,
	Phys.Lett. {\bf B282}, 50, (1992)
\bibitem{higgs} A.Aurilia, F.Legovini, E.Spallucci,
 Phys.Lett. {\bf B264}, 69, (1991)
\bibitem{stuck} S.Ansoldi, A.Aurilia,  E.Spallucci,
	``~Vacuum Energy Conversion into Dark Matter~'' Dept. of Theor.
	Phys.-- Univ. of Trieste preprint, in preparation
\bibitem{last}  P.Orland Nucl. Phys. B205 (1982) 107;\\
		M.C. Diamantini Phys.Lett. B388 (1996) 273\\
		F. Quevedo, C. Trugenberger
	        Nucl.Phys. B501 (1997) 143\\
	        S.Deser, A.Gomberoff, M.Henneaux, C.Teitelboim
		Nucl.Phys. B520 (1998) 179
		
		\end{thebibliography}
\end{document}